# THE MODIFIED MHD EQUATIONS

# (Revised Version)

**Evangelos Chaliasos**

365 Thebes Street

GR-12241 Aegaleo

Athens, Greece

*Abstract*

After finding the really self-consistent electromagnetic equations for a plasma, we proceed in a similar fashion to find how the MHD equations have to be modified accordingly. Substantially this is done by replacing the "Lorentz" force equation by the correct (in our case) force equation. Formally we have to use the vector potential instead of the magnetic field intensity. The appearance of the formulae presented is the one of classical vector analysis. We thus find a set of eight (8) equations in eight (8) unknowns, as previously known concerning the tradditional MHD equations.

**1. Introduction**

In a previous paper [2] we have found the equations of real electromagnetic self-consistency in a plasma. The two-fluid approximation was used, and we arrived to a set of 14 equations in 12 unknowns, as in the case of the Vlasov system of equations [7]. Substantially the "Lorentz" force had to be replaced by another force equation, consistent with the remaining equations. The approach was fully relativistic and the notation used was relativistic as well.

In the present paper the task is undertaken to do the same in the one-fluid approximation. This is the well-known MHD approximation, making use of the magnetic field intensity only (and not of the electric field intensity at all). We insert again the correct expression for the "Lorentz" force, that is the one consistent with the remaining equations. But now it is very convenient (if not obligatory) to use the vector potential instead of the magnetic field intensity itself. And, because the MHD approximation is not relativistic, we use conventional vector analysis notation. We will be able in this way to find a set of 8 equations in 8 unknowns, as in the case of tradditional MHD.

**2. Deduction of the equations**

As we know, the tradditional MHD equations are obtained from the equations of electromagnetism (and fluid dynamics) for a magnetic permeability $\mu = 1$ and an electrical conductivity $\sigma = \infty$. The MHD approximation is also non-relativistic, that is the MHD equations are vallid only for velocities $v/c \ll 1$. These same assumptions will be kept in our case as well.

First of all, since we will work with potentials (instead of field intensities), we have to choose a gauge. We will use as such the Coulomb gauge, that is

$$(\phi = 0 \ \&) \ \mathrm{div}\vec{A} = 0, \tag{1}$$

which is a special case of the Lorentz gauge previously used [2].

The equation for the force we will use is [2]

$$mc\frac{du^i}{ds} = -\frac{e}{c}A^i{}_{,k}u^k. \tag{2}$$

From this it easily results, in the case of a two-component plasma for simplicity, instead of the "Lorentz" force the expression

$$\vec{f} = -\frac{\rho_+}{c}\left[\frac{\partial \vec{A}}{\partial t} + (\vec{v}_+ \cdot \mathrm{grad})\vec{A}\right] + \frac{\rho_+}{c}\left[\frac{\partial \vec{A}}{\partial t} + (\vec{v}_- \cdot \mathrm{grad})\vec{A}\right], \tag{3}$$

or

$$\vec{f} = \frac{\rho_+}{c}\left[(\vec{v}_- - \vec{v}_+) \cdot \mathrm{grad}\right]\vec{A}, \tag{3'}$$

resulting in

$$\vec{f} = -\frac{1}{c}(\vec{j} \cdot \mathrm{grad})\vec{A}, \tag{3''}$$

where $\rho_+$ & $\rho_-$ (and $\vec{v}_+$ & $\vec{v}_-$) refer to ions & electrons respectively, with $\rho_+ + \rho_- = 0$ (and $\rho_+\vec{v}_+ + \rho_-\vec{v}_- = \vec{j}$).

Now, from the equation

$$\frac{\partial \vec{H}}{\partial t} - \mathrm{curl}(\vec{v}\times\vec{H}) = \frac{c^2}{4\pi\sigma\mu}\nabla^2\vec{H} \tag{4}$$

([6], (63.7)), we obtain in our case ($\sigma = +\infty$)

$$\frac{\partial \vec{H}}{\partial t} = \mathrm{curl}(\vec{v}\times\vec{H}). \tag{5}$$

Thus, because

$$\vec{H} = \mathrm{curl}\vec{A}, \tag{6}$$

we find

$$\mathrm{curl}\frac{\partial \vec{A}}{\partial t} = \mathrm{curl}(\vec{v}\times\mathrm{curl}\vec{A}), \tag{7}$$

so that

$$\frac{\partial \vec{A}}{\partial t} = \vec{v}\times\mathrm{curl}\vec{A} + \mathrm{grad}f_*, \tag{8}$$

where $f_*$ an arbitrary function. In order to determine $f_*$, we note that

$$\vec{E} = -\frac{1}{c}\vec{v}\times\vec{H} \tag{9}$$

([6], (65.9)), so that

$$\frac{\partial \vec{A}}{\partial t} = -c\vec{E} + grad f_*, \tag{10}$$

or

$$\vec{E} = -\frac{1}{c}\frac{\partial \vec{A}}{\partial t} + \frac{1}{c} grad f_*. \tag{11}$$

But

$$\vec{E} = -\frac{1}{c}\frac{\partial \vec{A}}{\partial t} - grad\phi \tag{12}$$

([4], (17.3)). Thus, since φ = 0 (Coulomb gauge, see (1)), it results in grad f* = 0, so that, from eqn. (8),

$$\frac{\partial \vec{A}}{\partial t} = \vec{v} \times curl\vec{A}. \tag{13}$$

This equation has to be taken as a Maxwell equation (one of the second pair), instead of ([4], (30.3)) with ∂**E**/∂t neglected ([6], p. 227)(cf. also [6], (65.2)).

From the Maxwell equations (and since the first pair reduces to identities in terms of the potentials), the following remains:

$$div\vec{E} = 0 \tag{14}$$

([4], (30.4)). But this is also an identity: because of the Coulomb gauge (eqns. (1)) we obtain from eqn. (12)

$$div\vec{E} = -\frac{1}{c}\frac{\partial}{\partial t} div\vec{A} = 0. \tag{15}$$

We also have the (mass) equation of continuity (coming from Fluid Mechanics), namely

$$\frac{\partial \rho}{\partial t} + div(\rho \vec{v}) = 0 \tag{18}$$

([6], (65.3)).

Thus, the equation of motion finally remains. It comes also from Fluid Mechanics, and is the Euler equation

$$\frac{\partial \vec{v}}{\partial t} + (\vec{v} \cdot grad)\vec{v} = -\frac{1}{\rho} grad P + \frac{\vec{f}}{\rho} \tag{19}$$

([5], (2.4)). Substituting **f** (instead of the "Lorentz" force) from (3"), we get

$$\frac{\partial \vec{v}}{\partial t} + (\vec{v} \cdot grad)\vec{v} = -\frac{1}{\rho} grad P - \frac{1}{\rho c}\left[(\vec{j} \cdot grad)\vec{A}\right], \tag{20}$$

But, because of (6), and the second pair's first Maxwell equation, written in the form

$$curl\vec{H} = \frac{4\pi}{c}\vec{j} \qquad \vec{j} = (c/4\pi) curl curl\vec{A}, \tag{21} \tag{23}$$

we can write for the force

$$\vec{f} = -(1/4\pi)\left\{\left[(curlcurl\vec{A}) \cdot grad\right]\vec{A}\right\}. \tag{24}$$

Thus finally (19) becomes

$$\frac{\partial \vec{v}}{\partial t} + (\vec{v} \cdot grad)\vec{v} = -(1/\rho)gradP - (1/4\pi\rho)\left\{\left[(curlcurl\vec{A}) \cdot grad\right]\vec{A}\right\}. \tag{25}$$

Thus, we have finally a set of 8 equations, namely (1) (1equation), (13) (3 equations), (18) (1 equation) and (25) (3 equations), in 8 unknowns, namely **A** (3 components), **v** (3 components), ρ (1 component) and P (1 component). Note that in tradditional MHD we have the same number for the equations and for the unknowns.

This set can be supplemented with the adiabatic equation (one additional unknown):

$$\frac{ds}{dt} \equiv \frac{\partial s}{\partial t} + \vec{v} \cdot grads = 0, \tag{26}$$

where s is the specific entropy (entropy per unit mass), ([6], (65.6); [5], (2.6); cf. also [5], (2.7)), or

$$\frac{d}{dt}\left(p\rho^{-\gamma}\right) = 0, \tag{27}$$

where γ is the adiabatic index (the ratio of the specific heats) ([3], p. 99; cf. also [1], p. 137), and an equation of state (one additional unknown too):

$$P = P(\rho, T) \tag{28}$$

([6], (65.5)). Thus we have totally 10 equations in 10 unknowns.

**Acknowledgement.** I wish here to express my sincere thanks, and deep gratitude, to Prof. E.N.Parker for protecting me from a serious error, which much affected equation (25).